\documentclass[aps,pra,twocolumn,superscriptaddress,longbibliography]{revtex4-1}
\usepackage{graphicx}
\usepackage{amsmath}
\usepackage{color}
\usepackage[colorlinks=true,citecolor=blue,linkcolor=magenta]{hyperref}
\hypersetup{pdfauthor = {Andrew Horsley},colorlinks = true, linkcolor = blue, urlcolor=blue, bookmarksnumbered =  true}

\begin{document}


\title{Microwave device characterization using a widefield diamond microscope}


\author{Andrew Horsley}
\email[]{andrew.horsley@unibas.ch}
\affiliation{Departement Physik, Universit\"{a}t Basel, CH-4056 Switzerland}
\affiliation{Laser Physics Centre, Research School of Physics and Engineering, Australian National University, 2601 Canberra, Australia}
\author{Patrick Appel}
\author{Janik Wolters}
\affiliation{Departement Physik, Universit\"{a}t Basel, CH-4056 Switzerland}
\author{Jocelyn Achard}
\author{Alexandre Tallaire}
\affiliation{Laboratoire des Sciences des Proc\'{e}d\'{e}s et des Mat\'{e}riaux (LSPM), CNRS, Universit\'{e} Paris 13, Sorbonne Paris Cit\'{e}, 99 avenue J.B. Cl\'{e}ment, 93430 Villetaneuse, France}
\author{Patrick Maletinsky}
\email[]{patrick.maletinsky@unibas.ch}
\author{Philipp Treutlein}
\affiliation{Departement Physik, Universit\"{a}t Basel, CH-4056 Switzerland}


\date{\today}

\begin{abstract}

Devices relying on microwave circuitry form a cornerstone of many classical and emerging quantum technologies. A capability to provide in-situ, noninvasive and direct imaging of the microwave fields above such devices would be a powerful tool for their function and failure analysis. In this work, we build on recent achievements in magnetometry using ensembles of nitrogen vacancy centers in diamond, to present a widefield microwave microscope with few-micron resolution over a millimeter-scale field of view, $130\,\mathrm{nT}\,\mathrm{Hz}^{-1/2}$ microwave amplitude sensitivity, a dynamic range of 48~dB, and sub-ms temporal resolution. We use our microscope to image the microwave field a few microns above a range of microwave circuitry components, and to characterize a novel atom chip design. Our results open the way to high-throughput characterization and debugging of complex, multi-component microwave devices, including real-time exploration of device operation.

\end{abstract}

\maketitle

Microwave (MW) devices play a critical role in telecommunications, defence, and quantum technologies. Device characterization via high resolution MW field imaging is a long-standing goal~\cite{Rosner2002,Sayil2005,Horsley2015thesis}, which promises to overcome the limitations of conventional characterization techniques. For example, it is difficult to identify internal features of complex devices using S-parameter measurements of reflection and transmission through external device ports~\cite{Deutschmann2006,Dubois2008}. A high-throughput MW imaging method would allow for fast prototype iteration, and for more adventurous development of novel device architectures. Furthermore, MW imaging is of interest for spin-wave imaging in magnonic systems~\cite{vanderSar2015,Du2017}, is under investigation for medical imaging~\cite{Nikolova2011,Chandra2015}, and can be used to characterize materials~\cite{Plassard2011} and biological samples~\cite{Tselev2016}. In recent years, alkali vapor cells with atoms in the ground~\cite{Boehi2012,Horsley2013,Affolderbach2015,Horsley2015,Horsley2016} or highly excited Rydberg~\cite{Fan2015a,Holloway2017,Wade2017} states, and nitrogen-vacancy (NV) centers in diamond~\cite{Wang2015,Appel2015,Shao2016a,Chang2017} have shown promise for intrinsically calibrated MW imaging in simple, vacuum- and cryogen-free environments. Ensembles of NVs in a widefield diamond microscope~\cite{Steinert2010,Glenn2015,Simpson2016,Glenn2017,Tetienne2017} provide an excellent balance between the sensitivity and wide field of view (FOV) offered by vapor cells and the nanoscale spatial resolution of single NV centers~\cite{Appel2016}, and so far have been primarily employed for imaging static and low-frequency magnetic fields. In this work, we demonstrate high-throughput widefield diamond microscopy for MW device characterization, enabled by a step-change we have achieved in microscope performance.

Our microscope integrates advances in camera speed, experiment control, novel diamond material, laser illumination, and the use of an intrinsically calibrated MW sensing scheme, to perform MW imaging with an unprecedented combination of temporal resolution, FOV and spatial resolution. We demonstrate a $\sim$0.5~mm$^2$ FOV with few-micron spatial resolution, a MW amplitude sensitivity of $130\,\mathrm{nT}\,\mathrm{Hz}^{-1/2}$, and a dynamic range of 48~dB. We have advanced the temporal resolution to the sub-ms regime, an order of magnitude beyond the previous state of the art~\cite{Shao2016a}, enabling dynamic probing of circuit operation, and real-time exploration of large-scale devices by scanning them under the microscope (Supplementary Movies~1, 2~\cite{Supplementary}). In addition, the accessible design of our microscope enables high-throughput measurements with rapid exchange of MW devices, demonstrating its applicability to industry-relevant environments.

\begin{figure*}
\includegraphics[width=1\textwidth]{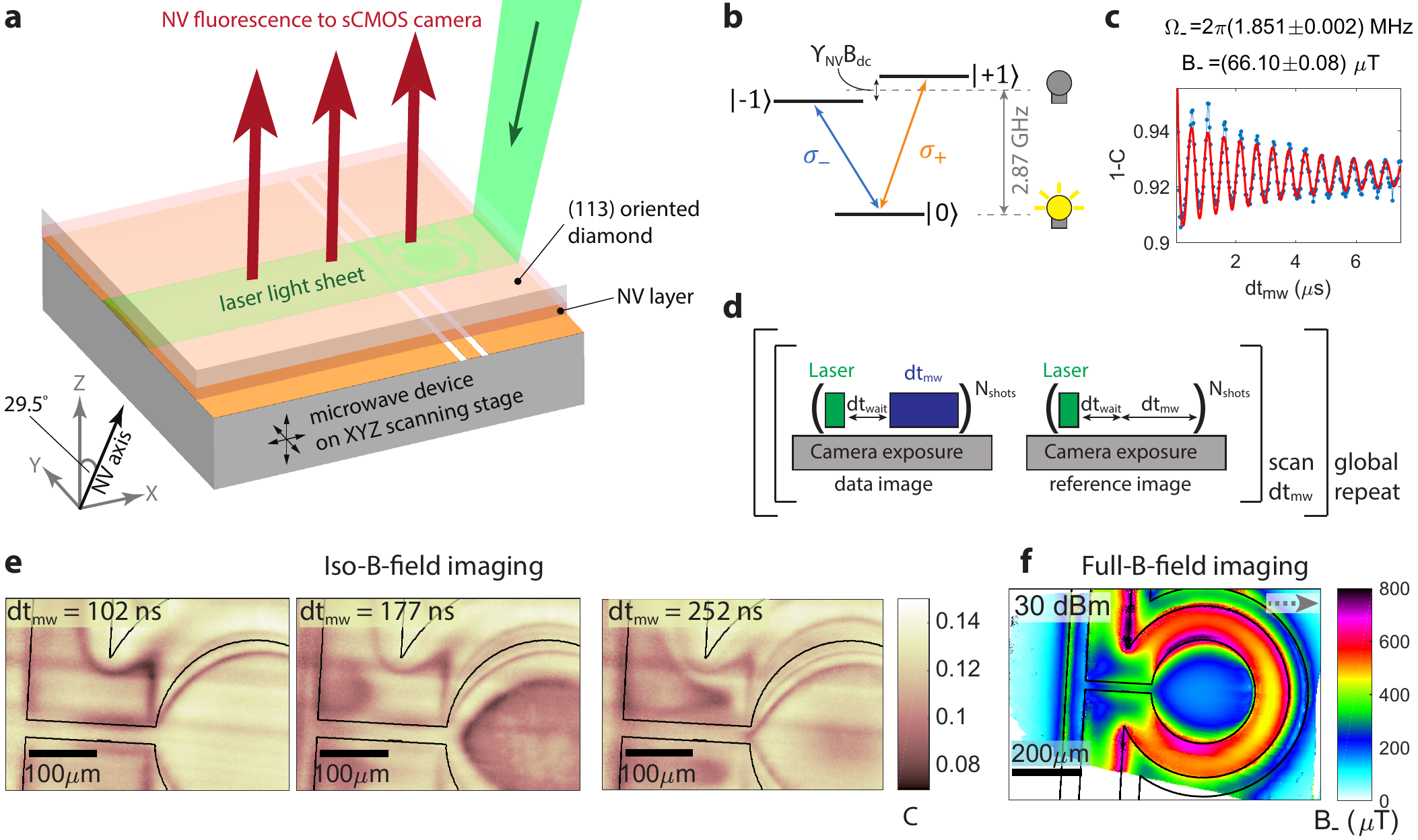}%
\caption{\label{fig:intro_figure} Widefield diamond microscope and MW imaging technique. \textbf{(a)} Schematic of the MW imaging setup. We performed imaging using NVs aligned along the (111) axis, tilted $29.5^{\circ}$ from the vertical in the XZ plane. \textbf{(b)} NV center ground state level structure, showing the $\sigma_\pm$ polarized MW transitions. \textbf{(c)} Rabi oscillations driven by a MW magnetic field between bright ($|0\rangle$) and dark ($|+1\rangle$ or $|-1\rangle$) fluorescing states. A fit using Eq.~\ref{eq:Rabi_fit} is shown in red. \textbf{(d)} Full-B-field imaging sequence. \textbf{(e)} Iso-B-field imaging mode, revealing the contours of the MW magnetic field. Images are shown for MW pulses of varying length (dt$_{\mathrm{mw}}$). Horizontal streaks are due to spatial variation in the 532~nm laser intensity. \textbf{(f)} Full-B-field image, obtained from a sequence of iso-B-field images where dt$_{\mathrm{mw}}$ is scanned, yielding a time-domain Rabi oscillation signal for each pixel (as shown in (c)). Pixel-wise fitting then yields an image of the MW magnetic field.}
\end{figure*}


\section{Microwave Imaging with the Widefield Diamond Microscope}

Our imaging goal is to determine the spatial profile of an inhomogeneous MW field of known frequency. We perform measurements by driving oscillations on an NV MW transition and measuring the oscillation frequency, which is proportional to the MW magnetic field amplitude. At the core of our microscope is a diamond containing a high-density layer of NV centers that can be brought in close proximity to a microwave device (Fig.~\ref{fig:intro_figure}a). In our particular setup, we use a (113) oriented diamond with a $\sim$$25\,\mu$m thick high density ($4\times10^{14}\,\mathrm{cm}^{-3}$) layer of NV centers preferentially oriented along the (111) axis, i.e. oriented at a $29.5^{\circ}$ angle from the diamond surface normal (see Supplementary Note~1~\cite{Supplementary})~\cite{Lesik2015}. NV centers are optically active lattice defects in diamond, with an electronic spin-1 ground state (Fig.~\ref{fig:intro_figure}b)~\cite{Doherty2013,Degen2017}. Excitation with 532~nm laser light stimulates state-dependent fluorescence, and pumps the NV population into the brightly fluorescing $|0\rangle$ state. To optimize the FOV for a given laser power, we excite the NVs using an in-plane sheet of 532~nm laser light (Fig.~\ref{fig:intro_figure}a, see Supplementary Note~1~\cite{Supplementary}). The $|0\rangle$ state is coupled to the darker $|\pm1\rangle$ states by MW magnetic dipole transitions. To detect MW fields, we first apply a dc magnetic field (B$_{\mathrm{dc}}$) to tune one of the ground state MW transitions into resonance with the MW frequency of interest. Using an appropriate pulse sequence (see below), we then measure the coherent Rabi oscillations driven by the MW between the coupled states, from which we extract the Rabi frequency (Fig.~\ref{fig:intro_figure}.c)~\cite{Boehi2010a,Wang2015}. The measurements resolve the microwave polarization, as each transition is sensitive to only a single polarization component of the MW magnetic field, B$_\pm$ for the respective $\sigma_\pm$ ($|0\rangle\rightarrow|\pm1\rangle$) transitions, with the polarization quantization axis parallel to the NV axis (see Supplementary Note~1~\cite{Supplementary}). A single NV axis can be used to measure both B$_+$ and B$_-$ at a given frequency, by reversing the B$_{\mathrm{dc}}$ direction in turn to tune each of the $\sigma_\pm$ transitions to the desired resonance frequency. Our measurements are intrinsically calibrated, as the Rabi oscillation frequency, $\Omega_\pm$, is related to the MW amplitude by $\mathrm{B}_\pm=\Omega_\pm/(2\pi\gamma_{\mathrm{NV}})$, through the well-characterized NV gyromagnetic ratio, $\gamma_{\mathrm{NV}}=28\,\mathrm{kHz}/\mu\mathrm{T}$.

We perform imaging by taking a series of data and reference images (Fig.~\ref{fig:intro_figure}.d). The data image sequence consists of a single (700~ns) laser pulse, followed by a single MW pulse input to the MW device under test (DUT). A wait time of $\mathrm{dt}_{\mathrm{wait}}=1.5\,\mu$s between the laser and MW pulses allows for relaxation of the optically excited NV electron through a metastable state. This sequence is repeated $\mathrm{N}_{\mathrm{shots}}\approx100$ times during a single camera exposure to accumulate fluorescence counts. The laser pulse both reads out the state of the NVs after the previous MW pulse, and prepares the NVs in the $|0\rangle$ state before the next MW pulse. We then take a reference image, with a sequence identical to the data image, but with the MW off. The data and reference images are combined pixel-wise to create a contrast image of the relative change in fluorescence induced by the microwave pulse, $\mathrm{C}=1-\mathrm{N}_{\mathrm{data}}/\mathrm{N}_{\mathrm{ref}}$, where $\mathrm{N}_{\mathrm{data}}$ ($\mathrm{N}_{\mathrm{ref}}$) is the fluorescence counts for a given data (reference) pixel. The use of a reference image reduces the sensitivity to spatial variation in fluorescence collection efficiency and noise (e.g. temporal laser intensity fluctuations) slower than the $\sim$ms separation of the data and reference images, limited by the camera frame rate. However, as we operate below the NV optical saturation level, spatial variation in laser intensity does result in variation in the initialization fidelity of NVs in the $|0\rangle$ state, and a proportional variation in contrast.

We operate in either of two imaging modalities: iso-B-field and full-B-field imaging~\cite{Boehi2010a,Appel2015}. Iso-B-field imaging (Fig.~\ref{fig:intro_figure}e) is a single-shot technique, providing the highest temporal resolution. We drive Rabi oscillations for a fixed duration, dt$_{\mathrm{mw}}$, leaving NV centers in a superposition of bright and dark states depending on the local B$_\pm$. Bright lines in the fluorescence contrast images occur at $\Omega_\pm=m\pi/\mathrm{dt}_{\mathrm{mw}}$, where $m$ is an integer, and therefore follow contour lines of B$_\pm$. Longer dt$_{\mathrm{mw}}$ pulses drive more Rabi oscillations, resulting in more closely spaced contour lines and a higher MW amplitude resolution. Calibration of the contour lines can be performed by counting them from a region far from the MW source (where B$_\pm\sim0$), inwards towards the MW source~\cite{Boehi2010a}. Although fast, iso-B-field images can be obscured by contrast variation due to e.g. inhomogeneous laser illumination (the cause of the horizontal stripes in Fig.~\ref{fig:intro_figure}e), and are limited in MW amplitude resolution by the optical imaging resolution and MW amplitude gradient~\cite{Appel2015}.

Full-B-field imaging (Fig.~\ref{fig:intro_figure}f) extends the iso-B-field technique by scanning dt$_{\mathrm{mw}}$. This gives a time-varying signal for each pixel in an image (Fig.~\ref{fig:intro_figure}c), from which we extract the Rabi frequency using the fit function
\begin{multline}\label{eq:Rabi_fit}
y=A-\Big(B\exp(-dt_{\mathrm{mw}}/\tau_{\mathrm{fast}})\\
+C\exp(-dt_{\mathrm{mw}}/\tau_{\mathrm{slow}})\Big)\sin(\Omega_\pm dt_{\mathrm{mw}}),
\end{multline}
where $A$, $B$, $C$, $\tau_{\mathrm{fast}}$, $\tau_{\mathrm{slow}}$ and $\Omega_\pm$ are fit parameters. Details of the fitting are discussed in Supplementary Note~2~\cite{Supplementary}. By directly measuring an oscillation frequency, our measurements become largely insensitive to spatial and temporal fluctuations in signal amplitude, and the B$_\pm$ amplitude resolution becomes limited only by the contrast detection noise. We demonstrate a 48~dB dynamic range in MW power (see Supplementary Note~1~\cite{Supplementary}), with the lower bound given by the NV coherence time in our diamond sample, and the upper bound given by the available MW power.

The microscope can be readily adjusted to optimize temporal resolution, dynamic range, field of view, sensitivity, or spatial resolution, as discussed in Supplementary Note~1~\cite{Supplementary}.
The DUT is mounted separately to the diamond, meaning that it can be scanned to build up composite images of arbitrary size. The MW images presented in this work (Figs~2-4) required 2-3~min of measurement time per image at a given DUT position. Separate DUT mounting and the open geometry of the microscope allow for high throughput MW device characterization, with DUT exchange and realignment a straightforward process that can be performed in under 10~min.

\begin{figure*}
\includegraphics[width=0.75\textwidth]{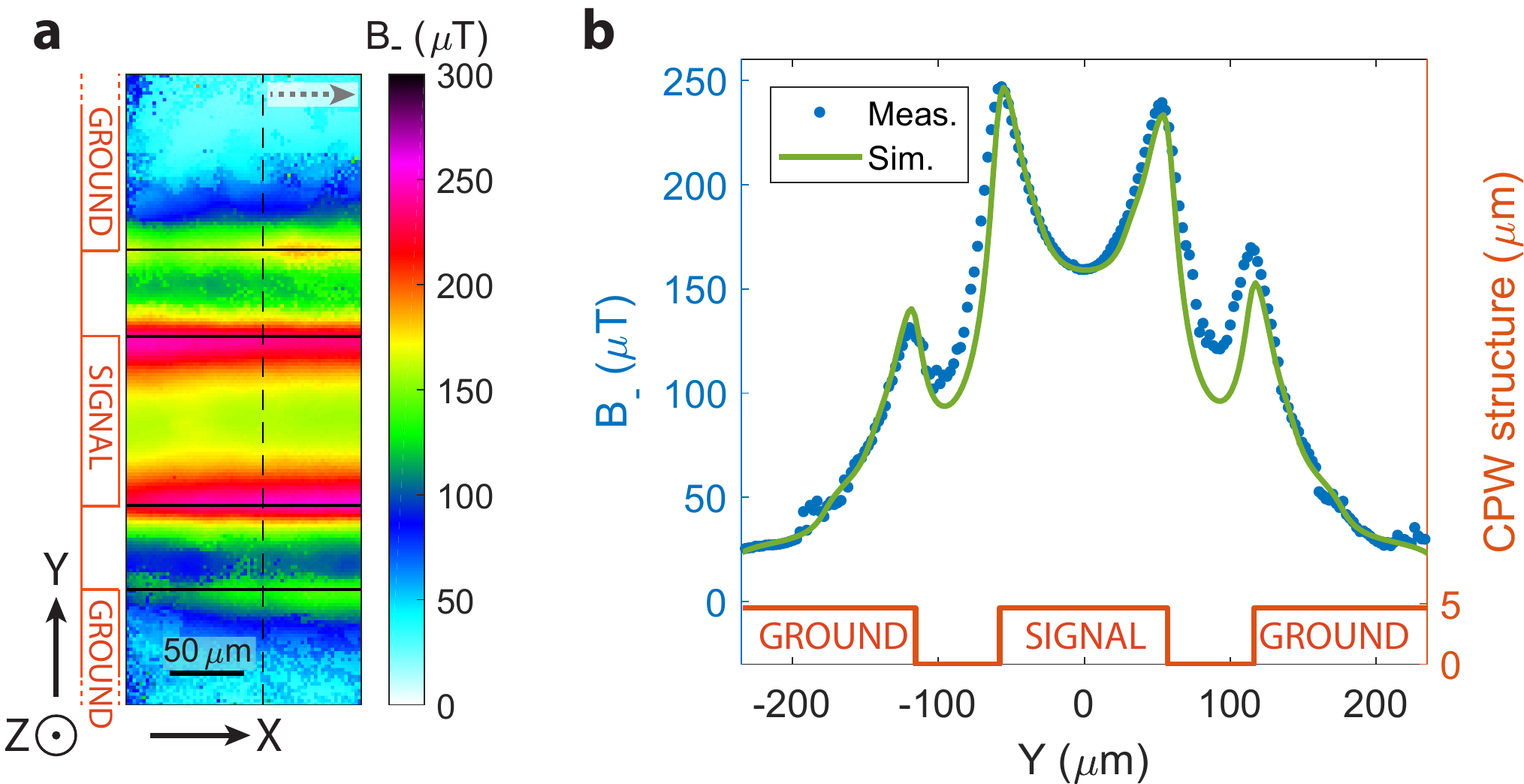}%
\caption{\label{fig:cpw_comparison} Validation measurement on a section of coplanar waveguide (CPW). \textbf{(a)} Measured MW field, B$_-$, with an input power to the chip of 22.6~dBm. Horizontal black lines outline the CPW structure. The NV axis is oriented $29.5^{\circ}$ from the Z axis, in the direction of the X axis, as indicated by the dashed grey arrow. \textbf{(b)} Line cut through the CPW (indicated by a dashed line in (a)), compared to simulation. }
\end{figure*}

\section{Microwave device characterization}

\begin{figure*}
\includegraphics[width=0.55\textwidth]{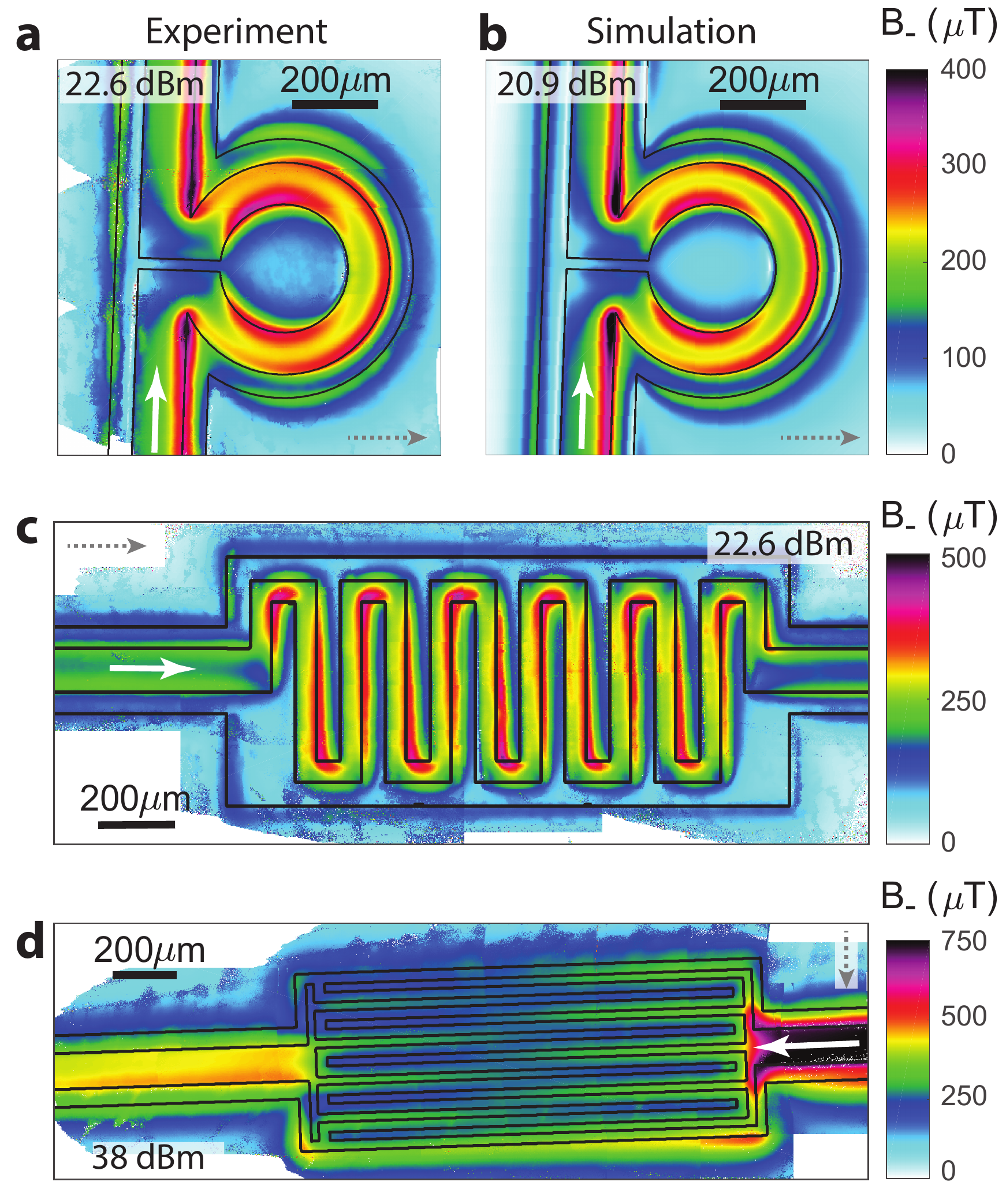}%
\caption{\label{fig:lumped_elements} Images of 2.77~GHz MW near-fields above various lumped-element structures. Measured \textbf{(a)} and simulated \textbf{(b)} field above an omega resonator. \textbf{(c)} Measured field above a meander line. \textbf{(d)} Measured field above an interdigital capacitor. Solid white arrows indicate the direction of MW propagation, dashed grey arrows indicate the projection of the NV axis onto the XY plane, and solid black lines show contours of the DUT structures. MW powers, given in dBm, are measured immediately before the input to the devices. The MW images for each structure are each stitched together from several smaller, overlapping images. }
\end{figure*}

We first validate our microwave imaging system by measuring the field above a section of coplanar waveguide (CPW) and comparing our results to simulation. The CPW has a $120\,\mu$m wide central signal line, with $54\,\mu$m gaps to ground planes on either side (see Supplementary Note~1~\cite{Supplementary}). Fig.~\ref{fig:cpw_comparison}a shows the measured B$_-$ component of the MW magnetic field at 2.77~GHz. We used the software Sonnet to simulate the 2D MW current distribution in the CPW (Supplementary Fig.~S5a~\cite{Supplementary}). We then calculated the resulting MW magnetic near field using the Biot-Savart law, and determined the B$_-$ projection onto the NV axis. To account for the finite sensing volume, we integrate the field over a range $Z=h\pm d/2$ above the CPW. The mean sensing height, $h$, was a fit parameter, whilst the sensing layer thickness, $d=14\,\mu$m, was given by the thickness of the laser light sheet (see Supplementary Note~1~\cite{Supplementary}). Fig.~\ref{fig:cpw_comparison}b compares the measured and simulated fields in a line cut across the CPW. We find good agreement with $h=12\,\mu$m and a microwave current in the signal line of $\mathrm{J}_{\mathrm{mw}}=50\,\mathrm{mA}$, without additional free fit parameters. Assuming the laser light sheet extended to the edge of the diamond, this corresponds to a chip-diamond separation of $h-d/2=5\,\mu$m. The simulated MW power is P$_{\mathrm{mw}}=\mathrm{J}_{\mathrm{mw}}^2\,\mathrm{Z}=20.9\,\mathrm{dBm}$, where $\mathrm{Z}\approx50\,\Omega$ is the waveguide impedance, corresponding to a 1.7~dB insertion loss of the measured $22.6\pm2$~dBm input power.

Figures~\ref{fig:lumped_elements}a,b show the measured and simulated B$_-$ component of the MW field above an omega loop resonator, a structure often used to produce uniform MW fields for control of quantum systems~\cite{Sasaki2016}. We see excellent agreement between the measured and simulated fields if we impose an asymmetric drive of the structure. That is, we performed the simulation using an asymmetric current distribution between the CPW signal line and each of the ground planes, with a current-imbalance of $\sim30\%$ between the left and right sides of the CPW signal line (see Supplementary Fig.~S5b~\cite{Supplementary}). By exploring the MW field in the CPW upstream from the omega loop (see Supplementary Movie~1~\cite{Supplementary}), we can gain an understanding of the origin of this asymmetric current.
Upstream of a $90^{\circ}$ bend in the CPW, the B$_-$ profile is symmetric (Fig.~S5c~\cite{Supplementary}), and is well-matched by fields produced by the same symmetric current distribution used in Fig.~\ref{fig:cpw_comparison}b. However, the MW field distribution after the bend (measured immediately before the omega loop) is best matched by fields produced by the asymmetric current distribution (Fig.~S5d~\cite{Supplementary}). We can therefore conclude that the asymmetric current is likely induced by either the CPW bend, or by reflections from any impedance mismatch between the omega loop and the CPW.
In addition to the field asymmetry produced by the MW current asymmetry, the B$_-$ asymmetry above the CPW downstream of the $90^{\circ}$ bend is also due the angle between the MW current and NV axis.

As a demonstration of the versatility of the microscope, we then imaged B$_-$  above an inductive meander line (Fig.~\ref{fig:lumped_elements}c) and an interdigital capacitor (Fig.~\ref{fig:lumped_elements}d). Like in the omega loop, we see that the MW field, and therefore current, is largely confined to edge modes, but avoids the outer edges of corners. In Fig.~\ref{fig:lumped_elements}c, the field features of the meander become blurred towards the left of the image, due to a $0.25^{\circ}$ tilt between the diamond and chip increasing the mean sensing distance from $h=11\,\mu$m on the right hand side of the image, to $h=20\,\mu$m on the left hand side. In Fig.~\ref{fig:lumped_elements}d, we see that the majority of the MW current is reflected at the CPW-capacitor interface. To increase the signal in the capacitor fingers, we input a relatively large P$_{\mathrm{mw}}=38\,\mathrm{dBm}$. The image shows that capacitive coupling does indeed transfer some MW current from the upstream to downstream fingers, however a significant current bypasses the capacitor fingers, flowing the through the ground planes surrounding the finger structure.

\begin{figure*}
\includegraphics[width=0.8\textwidth]{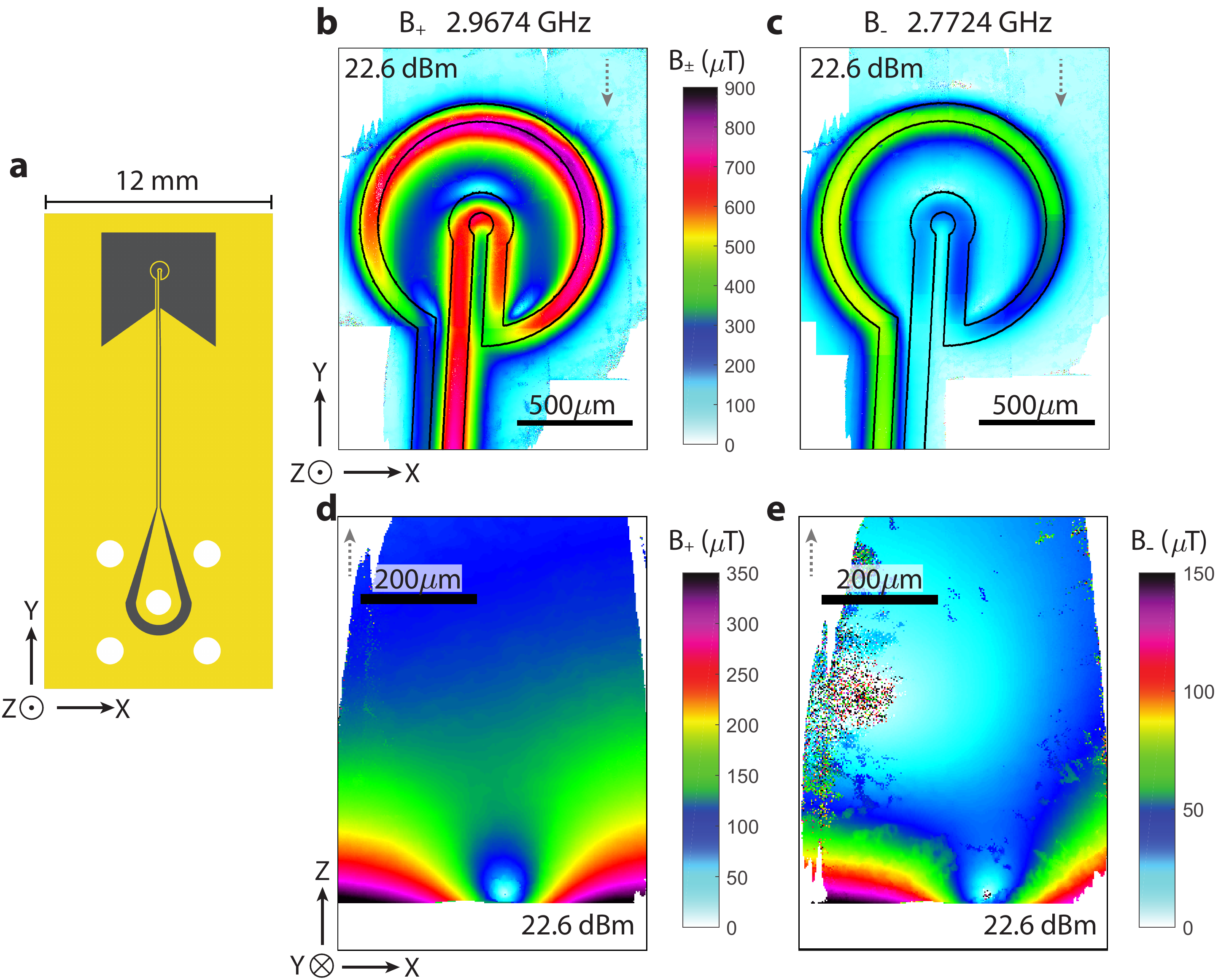}%
\caption{\label{fig:trap_chip} Prototype atom chip, producing an all-MW trapping potential. \textbf{(a)} Schematic of the trap chip, with metallization shown in yellow, etched surface in grey, and through-holes for an SMA jack in white. \textbf{(b,c)} MW field measured in the XY plane above the chip, at a MW frequency near a resonance of the chip ((b), 2.9674~GHz) and off-resonance ((c), 2.7724~GHz). \textbf{(d,e)} MW field measured in the XZ plane above the chip, near-resonance ((d), 2.9674~GHz) and off-resonance ((e), 2.7724~GHz). Dashed grey arrows indicate the projection of the NV axis onto the XY plane (b,c) or XZ plane (d,e), and solid black lines show contours of the trap chip structure. Images (b,c) are composite images stitched together from several smaller images, whilst (d,e) show a single imaging position. The particular mounting of the NV diamond limited the distance of approach to $80\,\mu$m above the chip surface in (c,e), as indicated by the white space below each MW image. }
\end{figure*}

In Fig.~\ref{fig:trap_chip}a, we present a prototype design of a novel atom chip, which is a type of device used for trapping and manipulating ultracold atoms, and a key platform for practical applications of quantum technologies that exploit such atoms~\cite{Reichel2011}. We use our microscope to perform an initial characterization, confirming that the chip provides an all-MW trapping potential, a new addition to the atom chip toolbox that will enable e.g. chip-generated trapping of atoms in hyperfine states untrappable using static fields~\cite{Spreeuw1994,Boehi2009}. The trapping potential, a minimum in the MW field above the chip, is created by counter-propagating MW currents through the inner and outer loops of a stripline.
The chip is a single-port device, with a CPW transitioning into the stripline used to form the trapping geometry, which then terminates to the ground plane.

We probe the chip at two frequencies, near and away from a resonance (identified using S-parameter measurements, see Supplementary Fig.~S6~\cite{Supplementary}). Near the resonance, at 2.9674~GHz, Fig.~\ref{fig:trap_chip}b shows the B$_+$ field in the XY plane above the trapping section. The MW field is reproducible in Sonnet simulations (Supplementary Fig.~S7~\cite{Supplementary}) by adjusting the relative impedances of the input and output ports at either end of the stripline (our simulations only considered the structure within the field of view of Fig.~\ref{fig:trap_chip}b) to produce a gradual attenuation of $|\mathrm{J}_{\mathrm{mw}}|$ through the loops. This smooth evolution of current implies that the resonance feature is not a resonance of the trapping loops, but is likely a resonance of the entire chip. In Fig.~\ref{fig:trap_chip}d we probe the chip away from the resonance, at 2.7724~GHz. This time, we image the B$_-$ component of the field, but due to the confinement of current flow in the stripline, we can reasonably expect that the MW field is close to linearly polarized, and so $|\mathrm{B}_{\mathrm{-}}| \approx|\mathrm{B}_{\mathrm{+}}|$ (see Supplementary Note~3 and Supplementary Fig.~S11 for measured example and discussion). As we might expect, the off-resonant field in Fig.~\ref{fig:trap_chip}d is weaker than the field on-resonance in Fig.~\ref{fig:trap_chip}b, with a factor of two reduction in the maximum measured field. A more surprising feature is the change in the spatial mode of the off-resonant field. We were unable to conclusively determine the cause of this mode change within the scope of this work.

The field distribution perpendicular to the chip, which we can image by mounting the chip perpendicular to the diamond, is particularly useful for characterizing the trapping potential, and Fig.~\ref{fig:trap_chip}d shows the B$_+$ component of a 2.9674~GHz MW field in the XZ plane. Note, however, that the change in chip-diamond orientation results in $\theta=90^{\circ}$ rotation of the NV axis with respect to the chip coordinate system, and a corresponding change in the definition of B$_+$. Figure~\ref{fig:trap_chip}d shows that a tight trapping potential is formed $100\,\mu$m above the chip, with a Z axis gradient of $3.7\,(2.2)\,\mu\mathrm{T}\,\mu\mathrm{m}^{-1}$ below (above) the trap, and a symmetric X axis gradient of $2.7\,\mu\mathrm{T}\,\mu\mathrm{m}^{-1}$.
We investigated the frequency dependence of the trapping field in Fig.~\ref{fig:trap_chip}e, where we image the B$_-$ component of a 2.7724~GHz field. We again see a trapping potential $100\,\mu$m above the chip. The field and trap gradients are 2-3 times lower than in Fig.~\ref{fig:trap_chip}d, matching the weaker fields seen off-resonance in Fig.~\ref{fig:trap_chip}c, and the trap position is shifted by $\Delta \mathrm{X} = +40\,\mu$m, qualitatively matching the asymmetry along X in Fig.~\ref{fig:trap_chip}c compared to Fig.~\ref{fig:trap_chip}b. In future versions of the chip, it may be possible to take advantage of this to scan the trap position by changing the input MW frequency (adjusting both the atoms' resonant frequency with a dc magnetic field, and P$_{\mathrm{mw}}$ as necessary). A second, broad minimum also appears in Fig.~\ref{fig:trap_chip}e, $430\,\mu$m above the chip surface, whose origin is unclear.

The surprising features of the measured microwave field distributions on the structures in Figs.~\ref{fig:lumped_elements} and \ref{fig:trap_chip} underline the importance of performing high-resolution microwave field measurements. Our imaging method is ideally suited for this task, as it provides calibrated images, which are well reproduced by simulations on simple structures such as the coplanar waveguide in Fig.~\ref{fig:cpw_comparison}. The detailed interpretation of the observed field distributions on more complex circuits such as in Fig.~\ref{fig:lumped_elements} and \ref{fig:trap_chip} will require further dedicated studies of microwave propagation on these structures.

\begin{figure*}
\includegraphics[width=0.75\textwidth]{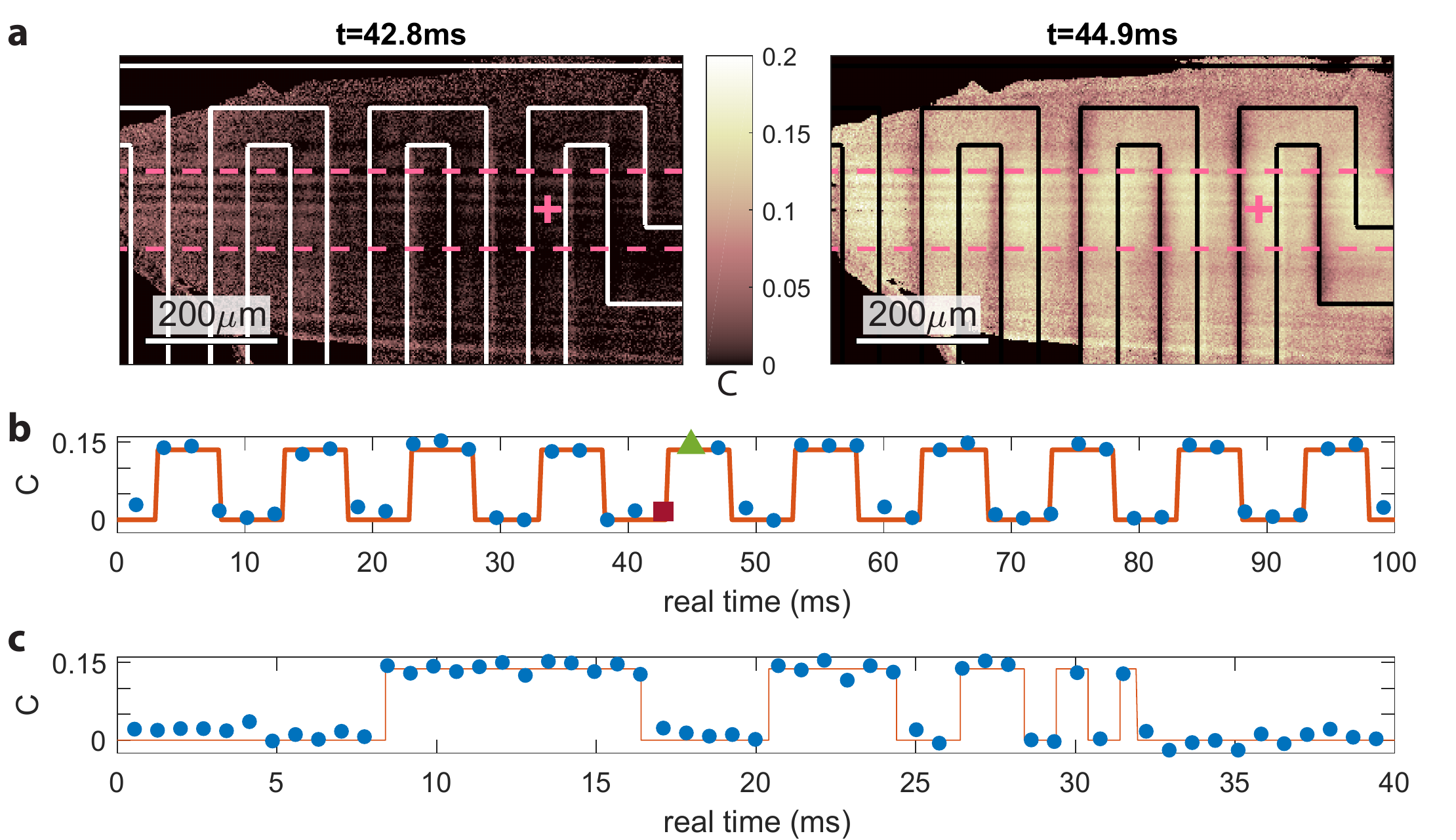}%
\caption{\label{fig:TemporalResMeander} Demonstration of temporal resolution in the iso-B-field imaging mode. \textbf{(a)} Iso-B-field images of B$_-$ above a section of the meander line chip, with a 2.77~GHz MW signal off (left) and on (right). The full vertical field of view is 200~pixel rows. Horizontal dashed lines indicate the field of view for 50~rows. The meander structure is outlined in white. \textbf{(b)} Time trace for the pixel denoted by a pink cross in (a), as a MW signal is pulsed on and off. The time steps corresponding to the example images are indicated by a red square (MW off) and a green triangle (MW on). \textbf{(c)}  Time trace with 50 pixel rows read out, showing sub-ms temporal resolution. }
\end{figure*}

\section{Temporal resolution}

We achieve our highest temporal resolution in the iso-B-field imaging mode, with reference images taken offline. The temporal resolution in our current system is ultimately limited by the camera frame rate, which is proportional to the number of pixel rows read out. Reading out 360~columns and 200~rows, corresponding to a $850\times470\,\mu\mathrm{m}^2$ field of view, and choosing $\mathrm{N}_{\mathrm{shots}}=50$ shots/exposure, we achieve a temporal resolution of 2.2~ms. We demonstrate this $\sim$kHz frame rate in Figs.~\ref{fig:TemporalResMeander}a,b, where we measure a pulse train with a 5~ms on/off cycle. Each pixel in Fig.~\ref{fig:TemporalResMeander}a provides a similar time-trace signal to that shown in Fig.~\ref{fig:TemporalResMeander}b. To explore our ability to detect externally controlled signals, we used an independent pulse generator to switch a P$_{\mathrm{mw}}=22.6\,\mathrm{dBm}$, 2.77~GHz signal. The imaging sequence was identical to the reference image sequence in Fig.~\ref{fig:intro_figure}d, but with a fixed gap between laser pulses of $\mathrm{dt}_{\mathrm{wait}}+\mathrm{dt}_{\mathrm{mw}}=530\,$ns. Reducing the readout to 50~rows further improves the temporal resolution to 0.7~ms, as shown in Fig.~\ref{fig:TemporalResMeander}c, where we measure a train of pulses between 8~ms and 0.5~ms in length. The measurements in Fig.~\ref{fig:TemporalResMeander} used a short sensing time, of dt$_{\mathrm{mw}}=30\,\mathrm{ns}$. However, longer sensing periods (e.g. 1$\sim$10$\,\mu$s), which would provide images with richer contour lines, can be used without loss of temporal resolution, as the frame rate was dominated by camera readout.

\begin{figure*}
\includegraphics[width=1\textwidth]{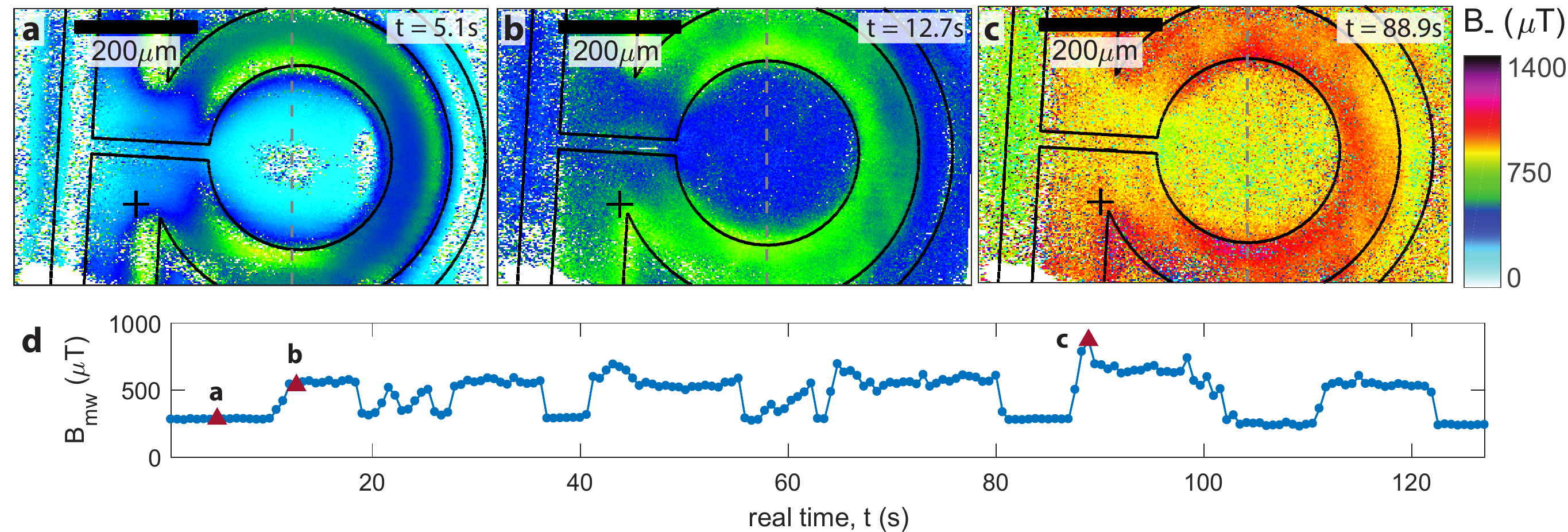}%
\caption{\label{fig:TemporalResOmega} Full-B-field MW imaging at a 1.6~Hz rate. \textbf{(a-c)} Frames from Supplementary Movie~2~\cite{Supplementary}, showing the B$_-$ component of a 2.77~GHz MW field above the omega loop as the pins of the device input port are shorted to one-another. (a) corresponds to the unperturbed device. MW field profiles along the vertical grey dashed lines are given in Supplementary Fig.~S8~\cite{Supplementary}. \textbf{(d)} Time trace for the pixel indicated by black crosses in the images in (a-c). Red triangles indicate the times of the frames shown in (a-c). }
\end{figure*}

To combine full-B-field imaging with high bandwidth, we performed fast scans of dt$_{\mathrm{mw}}$, with an experiment sequence identical to that used for the images in Figs.~\ref{fig:lumped_elements},\ref{fig:trap_chip}, but with only a single global repeat. There is a tradeoff between temporal resolution and dynamic range, through the choice of number of dt$_{\mathrm{mw}}$ steps. We chose a 100-timestep sequence, and performed full-B-field B$_-$ imaging with a 1.6~Hz frame rate, using $\mathrm{N}_{\mathrm{shots}}=195$ shots/exposure, 200~pixel rows, and with reference images taken online. This imaging mode allows for an intuitive exploration of devices and microwave near fields, as demonstrated in Supplementary Movie~1~\cite{Supplementary}, where we scan the omega loop chip to explore MW fields upstream of the loop. Figure~\ref{fig:TemporalResOmega} shows frames and a pixel time-trace from a second movie taken of the omega loop chip (Supplementary Movie~2~\cite{Supplementary}), where we deliberately perturbed the microwave current modes by periodically shorting the signal and ground pins of the MW input port. We see the MW amplitude and spatial mode change dramatically, as we change the spatial mode and efficiency of microwave launch into the device. Figures~\ref{fig:TemporalResOmega}b,c show improved MW field homogeneity and strength in the center of the omega loop, compared to the unperturbed case in Fig.~\ref{fig:TemporalResOmega}a (see also profiles of the fields in Supplementary Fig.~S8~\cite{Supplementary}). This highlights a perhaps rarely considered degree of freedom in experimentally optimising devices, where one could envisage, for example, optimising a device for manipulating quantum systems by shorting the input port pins in a controlled manner.

\section{Conclusions and Outlook}

We have presented a widefield diamond microscope based on NV centers, and demonstrated its substantial potential for characterizing MW devices. Our sub-millisecond temporal resolution represents an order of magnitude improvement over the previous $\sim$10~ms state of the art for diamond microscopes. Our microscope provides intrinsically calibrated, polarization-resolved MW field images, with $130\,\mathrm{nT}\,\mathrm{Hz}^{-1/2}$ sensitivity, 48~dB dynamic range, and few-micron spatial resolution over a $\sim0.5\,\mathrm{mm}^2$ FOV.
In this work, we have performed MW field imaging around 2.77-2.97~GHz, however larger dc magnetic fields can be applied to tune the imaging frequency up to hundreds of GHz~\cite{Horsley2016, Stepanov2015,Aslam2015}.
With the addition of a control MW field to perform dynamic decoupling, weak MW fields could be detected up to the $1/\mathrm{T}_2$ limit~\cite{Joas2017,Stark2017}. With delta-doped NV surface layers, the spatial resolution could be further extended to the sub-micron range~\cite{Steinert2010}, or superresolution techniques could be employed to provide resolution down to the nanoscale~\cite{Chen2013,Arai2015}.
We have also identified ways to improve our fluorescence count rate, and thus signal to noise ratio, by over 5~orders of magnitude (see Supplementary Note~1~\cite{Supplementary}).

Potential further applications of our high temporal resolution microscope include studies of MW device failure, changes in MW device operation with duty cycle, and monitoring integrated circuitry activity, e.g. logic processes on processor chips~\cite{Nowodzinski2015,Shao2016a}. With the addition of a dc magnetic field gradient, our microscope could operate as an analogue MW spectrum analyser~\cite{Chipaux2015}, providing feedback on sub-ms timescales. MW electric field imaging could be explored, detecting the electric field via Rabi oscillations driven on the electric transition coupling the $|\pm1\rangle$ states~\cite{Barfuss2015}, with a dc magnetic field used to tune the $|\pm1\rangle$ state splitting to resonance with the MW field.
Widefield diamond microscopes hold promise for extending sub-micron widefield microscopy to MW medical imaging~\cite{Nikolova2011,Chandra2015,Tselev2016}, which could enable new applications on the cellular level. With the use of a control MW field, materials analysis using scanning MW microscopy~\cite{Plassard2011} could be extended to widefield imaging applications.

In addition to MW field imaging, there are a number of further modalities in which our microscope can be employed, including imaging spin waves in solid state systems~\cite{vanderSar2015,Du2017} to investigate e.g. spin transport waveguides based on ferromagnetic domain walls~\cite{GarciaSanchez2015} and as a possible alternative to Brillouin light scattering techniques~\cite{Sebastian2015}. Using our 0.7~ms resolution, 2-point differential dc magnetic field imaging can be performed at a 700~Hz rate. This is of interest e.g. for monitoring biological processes such as neuron firing~\cite{Hall2012} and magnetic nanoparticle formation~\cite{LeSage2013}, and for microfluidics applications~\cite{Yao2014,Issadore2014}.

\textbf{Acknowledgments:} This work was supported by the Swiss National Science Foundation (SNFS) and DIADEMS. We thank B. Shields, J. Wood, L. Thiel, and M. Ganzhorn for laboratory help and general advice on NV centers, A. Edmonds and M. Markham for providing the (113)-oriented substrates used for CVD overgrowth, G. Witz and T. Julou for help with Micro-Manager, and M. W. Doherty for careful reading of the manuscript.

\textbf{Author contributions:} A.H., P.M. and P.T. conceived the project idea. A.H. designed, built, and operated the microscope, with input from P.A. and J.W.. P.A. and A.H. validated experiment measurements with simulations. J.A. and A.T. fabricated the diamond sample. A.H. and P.T. designed the prototype atom chip and other microwave devices. P.M. and P.T. supervised the project. A.H. wrote the manuscript, with input and discussion from all authors.

\textbf{Supplementary references:} References~\cite{Barson2017, Roth1989, Budker2007, Stuurman2010, Edelstein2014, Clevenson2015, Hanson2008, Dreau2011, Mildren2013, Vanier1989} are cited in the Supplementary Information~\cite{Supplementary}.

\bibliography{bibliography_NV_widefield}

\end{document}